\documentclass[journal]{IEEEtran}

\usepackage{cite}
\usepackage{amsmath}
\usepackage{array}
\usepackage{graphicx}
\usepackage{bm}
\usepackage[dvipsnames]{xcolor}
\usepackage{float}
\usepackage[hyphens,spaces,obeyspaces]{url}
\usepackage{makecell}
\usepackage{MnSymbol}
\usepackage{adjustbox}
\usepackage{booktabs}
\usepackage{nccmath}

\begin{document}

\title{Analysis of Electrostatic MEMS Using Energy-Charge Landscape}

\author{Raghuram Tattamangalam Raman, Arvind Ajoy and Revathy Padmanabhan  \thanks{R. Tattamangalam Raman, A. Ajoy and R. Padmanabhan are with Electrical Engineering, Indian Institute  of Technology
Palakkad, Palakkad, India. e-mail: 121704004@smail.iitpkd.ac.in; arvindajoy@iitpkd.ac.in; revathyp@iitpkd.ac.in}
}

\maketitle

\begin{abstract}
A common way to analyze electrostatic Micro-Electro-Mechanical Systems (MEMS) actuators is to use their energy-displacement landscape. Here, we describe an alternative approach to analyze electrostatic MEMS actuators using their energy-charge landscape. This technique involves coordinate transformation from displacement to charge, thereby formulating the Hamiltonian of electrostatic MEMS actuators in terms of charge. We investigate the use of the energy-charge landscape to analyze static pull-in, dynamic pull-in and pull-out phenomena. The voltage expressions derived using this method are identical with those derived using the conventional energy-displacement landscape. In addition, we also obtain the expressions for charge under static and dynamic pull-in conditions. This work can aid in the design and analysis of electrostatic MEMS devices. As a case study, the analysis of a feedback capacitor-MEMS actuator system is presented to illustrate the application of the energy-charge landscape.
\end{abstract}


\IEEEpeerreviewmaketitle

\section{Introduction}
\IEEEPARstart{E}{lectrostatic} Micro-Electro-Mechanical Systems (MEMS) actuators form the backbone of a wide range of devices such as accelerometers, MEMS switches, display devices etc \cite{choudhary2016mems,gad2005mems,rebeiz2004rf}. The popularity of these devices is driven by the fact that electrostatic actuation is highly energy efficient. Electrostatic actuators primarily involve coupling between mechanical and electrical domains. The energy-landscape is a convenient method to analyze phenomena like static pull-in, dynamic pull-in and stability of electrostatic MEMS \cite{marques2005modelling,fargas2005electrostatic,fargas2007resonant,marques2016energy,nemirovsky2001methodology}.  In most textbooks on electrostatic MEMS actuators \cite{senturia2007microsystem,madou2002fundamentals,allen2005micro,vinoy2014micro,lee2011principles,younis2011mems},  the working of the system is described in terms of displacement of the movable part. Phase plane analysis  \cite{kalyanaraman2005nonlinear,younis2011mems} also uses displacement to describe the dynamics and stability of the actuator. In this article, we present the analysis of electrostatic MEMS actuators with charge as the parameter to describe their statics and dynamics. Using the energy-charge landscape, we derive expressions for voltage and charge under static pull-in, dynamic pull-in and pull-out conditions. 

Why is analysis based on energy-charge landscape relevant? To address this question, we look at some examples where the actuator analysis involves charge. Electrostatic actuation driven by voltage suffers from pull-in instabilities, wherein the mechanical restoring force cannot balance the electrostatic force beyond a certain limit. For example, a typical MEMS cantilever experiences static pull-in at one-third \cite{younis2011mems} of the air-gap. Various techniques and control strategies are employed to modify the pull-in regime in electrostatic MEMS actuators -- for instance, the pull-in instability in electrostatic MEMS devices can be avoided by connecting a feedback capacitor in series \cite{seeger1997stabilization,chan2000electrostatic} with the MEMS devices. A Metal--Oxide--Semiconductor (MOS) capacitor operating in depletion mode,\cite{seeger1997stabilization} connected in series, can also stabilize electrostatically actuated devices. The pull-in limit can also be improved by using a memristor \cite{almeida2015mems} as a feedback sensing element. Negative capacitance using a ferroelectric capacitor connected in series \cite{masuduzzaman2014effective, raghu2019dynamic} can modulate the pull-in regime as well. In all the aforementioned examples, the working of the system can conveniently be investigated using charge. Thus, it is relevant to analyze electrostatic MEMS based on the energy-charge landscape. Ref. \cite{masuduzzaman2014effective} uses an energy-charge based approach to analyze the static response of the ferroelectric negative capacitance-electrostatic MEMS hybrid actuator. However, the energy profile of the MEMS actuator used therein, is valid only at points of static equilibrium. Hence, this cannot be used to understand the dynamics of MEMS actuators (we discuss this in more detail in Section III). Our goal is to showcase an energy-charge based approach to investigate electrostatic MEMS that addresses both statics and dynamics. We achieve this by employing a coordinate transformation from displacement to charge, in the Hamiltonian formalism. In the presence of damping, the proposed method also allows us to estimate parameters such as air-gap and spring constant. Given the importance of charge in MEMS applications, this method will contribute to its analysis and design. 

This paper is organized as follows. Section II reviews the statics and dynamics of an electrostatic MEMS actuator. Section III presents the Hamiltonian formalism based on coordinate transformation. Section IV describes the analysis of the MEMS actuator based on the energy-charge landscape. Section V presents the impact of damping. Section VI addresses the scope and limitations of the proposed method. Section VII presents a case study, showcasing the use of the energy-charge landscape in the analysis of a feedback capacitor-MEMS actuator system. Finally, section VIII presents our conclusions.

\section{\label{sec:Hamil_EX}Review of Statics and Dynamics of An Electrostatic MEMS Actuator}
\begin{figure}[t]
    \centering
    \includegraphics[scale=1.0]{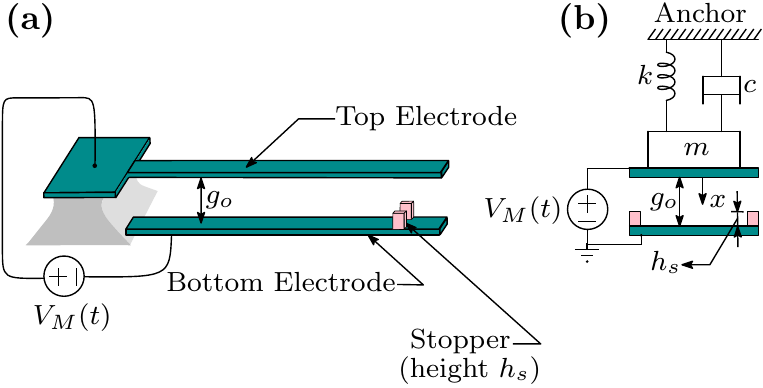}
    \caption{(a) Schematic of an electrostatic MEMS cantilever type actuator. (b) Equivalent 1-DOF
model with parameters: mass $m$, spring constant $k$, damping coefficient $c$, air-gap $g_o$ and displacement $x$.}
    \label{fig:MEMS_1DOF}
\end{figure}

In this section, the electro-mechanical response of an electrostatic MEMS cantilever type actuator excited by a voltage source is analyzed. We use a one degree-of-freedom (1-DOF) model, as depicted in Fig. \ref{fig:MEMS_1DOF}, to represent the electrostatic MEMS actuator. This is a lumped parameter model that approximates the MEMS actuator as a variable parallel plate capacitor, consisting of a fixed bottom electrode and a movable top electrode separated by an air-gap $g_o$. The inertia, energy dissipation, and stiffness of the device are modeled using a mass $m$, a viscous damper with damping coefficient $c$, and a spring of spring constant $k$, respectively. This lumped parameter model is a simplified representation \cite{ijntema1992static,senturia2007microsystem,younis2011mems,lee2011principles} that can be used to analyze the statics and dynamics of the system. Modal analysis indicates that the effective mass is less than the actual mass of the electrode \cite{younis2011mems,rebeiz2004rf}. However, we assume that the effective mass is equal to the actual mass of the movable electrode \cite{marques2016energy,qian2017sub}. Note that this does not change the essence of the analysis presented here. The excitation is denoted by an input voltage $V_M(t)$, where $t$ denotes time.       
\begin{table}[b]
\caption{\label{tab:Table1}%
Parameters of the MEMS actuator used in this work
}
	\centering
	\begin{tabular}{ll}
		\toprule
		\textbf{Parameter} & \textbf{Value}\\
		\midrule
		Length of the cantilever, $L$ & $160 ~\mu m$ \\
		Width of the cantilever, $W$ & $6 ~\mu m$ \\
		Thickness of the cantilever, $T$ & $2 ~\mu m$ \\
		Area, $A_M$ = $L$ $\times$ $W$ & $960 \times 10^{-12}~m^2$\\
		Cantilever Material & Silicon (Si)\\
		Young's Modulus, $E$ & $150 ~GPa$~\cite{kim2001111} \\
		Density, $D$ & $2330 ~kg/m^3$~\cite{shackelford2016crc} \\
		Mass, $m = D \times A_M \times T$ & $4.4736 \times 10^{-12}  ~kg$\\
		Spring Constant, $k = \frac{E~W~T^3}{4~L^3}$ & 0.439 $N/m$~\cite{senturia2007microsystem}\\
		Initial air-gap, $g_o$ & $3~\mu m$\\
		Stopper height, $h_s$ & $1.4~\mu m$\\
		Permittivity of free space, $\epsilon_o$ & $8.854 \times 10^{-12}~F/m$\\
		\bottomrule
	\end{tabular}
\end{table}	
The displacement of the top electrode, denoted by the dynamical variable $x$, is limited by means of a pair of stoppers of height $h_s$. The stoppers are made of insulating material and prevent shorting the top and bottom electrodes \cite{iannacci2009measurement,giacomzzi2013rf,autizi2011reliability}. These stoppers minimize the area of contact when the top electrode snaps down on to the bottom electrode and thus reduce the effect of surface forces. Keeping this in mind, we neglect the effect of surface forces in our analysis. For ease of analysis, damping coefficient $c$ is set to zero; we consider the case of non-zero damping later in Section V. The parameters of the electrostatic MEMS actuator used in this work are listed in Table \ref{tab:Table1}. The dimensions listed are fairly typical for MEMS cantilevers \cite{leus2008dynamic,shekhar2012switching,o2001mems}.
\begin{figure}[t]
    \centering
    \includegraphics[scale=1.0]{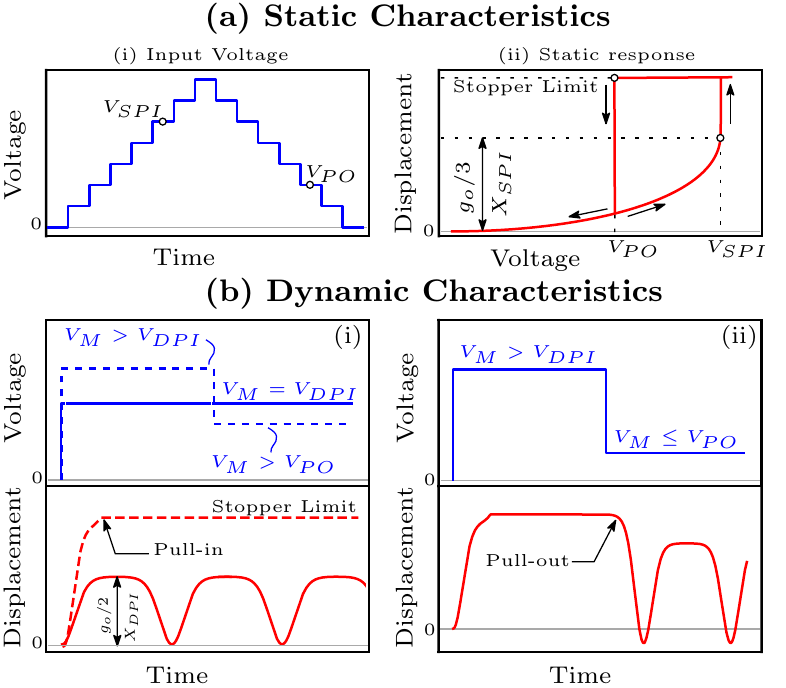}
    \caption{(a) Typical static characteristics depicting pull-in and pull-out. (b) Typical dynamic characteristics depicting pull-in and pull-out (without damping). Applied step-input and  corresponding actuator response: (i) before dynamic pull-in (solid line); after dynamic pull-in and without pull-out (dashed line), (ii) after dynamic pull-in and with pull-out. }
    \label{fig:Input_Output}
\end{figure}
\begin{table}[b]
\caption{\label{tab:Table2}%
Pull-in and pull-out of an electrostatic MEMS actuator. Values correspond to parameters listed in Table \ref{tab:Table1}.
}
	\centering
	\begin{adjustbox}{width=\columnwidth,center}
	\begin{tabular}{lcc}
		\toprule
		\textbf{Parameter} &\textbf{\makecell[l]{Expression$^\#$}}& \textbf{Value}\\
		\midrule
		\makecell[l]{Static pull-in voltage, $V_{SPI}$} & $\sqrt{(8~k~g_o^3)/(27~\epsilon_o~A_M)}$ & 20.33 $V$\\
		%\noalign{\vskip 1mm}    

		\makecell[l]{Static travel range, $X_{SPI}$} & $g_o/3$ & 1 $\mu m$\\
		%\noalign{\vskip 1mm}    

		\makecell[l]{Dynamic pull-in voltage, $V_{DPI}$} & $\sqrt{(k~g_o^3)/(4~\epsilon_o~A_M)}$ & 18.67 $V$\\
		%\noalign{\vskip 1mm}    

		\makecell[l]{Dynamic pull-in displacement, $X_{DPI}$} & $g_o/2$ & 1.5 $\mu m$\\
		%\noalign{\vskip 1mm}    

		\makecell[l]{Pull-out voltage, $V_{PO}$} & $\sqrt{(2~k~h_s^2~(g_o - h_s))/(\epsilon_o~A_M)}$ & 18 $V$\\
		\bottomrule
		\footnotesize{$\#$ From Refs.\cite{younis2011mems,fargas2005electrostatic}}
	\end{tabular}
	\end{adjustbox}
\end{table}

The typical static and dynamic characteristics of an electrostatic MEMS actuator, without damping \cite{raghu2019dynamic}, are illustrated in Fig. \ref{fig:Input_Output}. The static response, obtained by the application of a slowly varying input, shows a static pull-in voltage $V_{SPI}$, beyond which the top electrode snaps down, resulting in static pull-in. The maximum distance in the air-gap upto which the actuator can attain stable equilibrium is called the travel range $X_{SPI}$ \cite{younis2011mems}. The dynamic response of the actuator is characterized by applying a step-input of amplitude $V_M$. The response of the actuator, in the absence of damping, is oscillatory, for $V_M$ less than the dynamic pull-in voltage $V_{DPI}$. The maximum value of this oscillatory displacement is called dynamic pull-in displacement $X_{DPI}$ \cite{younis2011mems}. For any $V_M>V_{DPI}$, the top electrode snaps down, resulting in dynamic pull-in. After achieving pull-in (static or dynamic), the top electrode gets detached from the bottom electrode when the input is less than or equal to the pull-out voltage  $V_{PO}$, thereby resulting in pull-out \cite{younis2011mems}. After pull-out, the response of the actuator (in the absence of damping) is oscillatory, as shown in Fig. \ref{fig:Input_Output}(b)(ii). The expressions for the voltage and displacement, and their corresponding values for the designed MEMS actuator are summarized in Table \ref{tab:Table2}.
\begin{figure*}[t]
    \centering
    \includegraphics[scale=1.0]{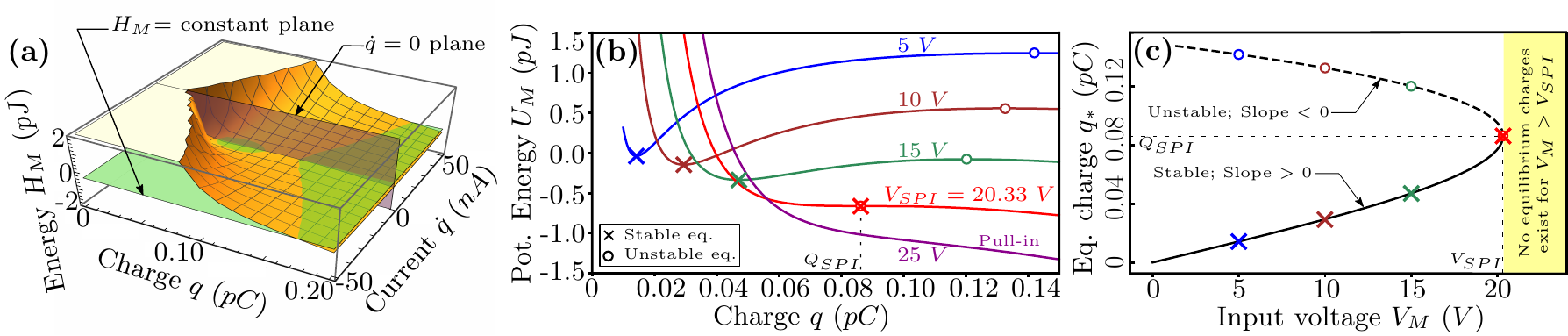}
    \caption{(a) Total energy ($H_M$) plotted as a function of charge ($q$) and current ($\dot{q}$) for an input voltage $V_M = 18 ~V$. Projection on the plane $\dot{q} = 0$ gives the potential energy-charge plot. Projection on the plane $H_M$ = constant gives the phase plane plot, as shown in Fig. \ref{fig:Dynamic_EQ_All}(d). (b) Potential energy ($U_M$) - charge ($q$) plot for different input voltages. The stable and unstable equilibrium charges coincide at the static pull-in charge $Q_{SPI}$, where input voltage $V_M$ equals static pull-in voltage $V_{SPI}$. (c) Equilibrium charge ($q_*$) - input voltage ($V_M$) plot. The stable (unstable) equilibrium charges lie on the plot where the slope is positive (negative). No equilibrium charges exist for $V_M>V_{SPI}$, resulting in static pull-in.}
    \label{fig:MEMS_3D}
\end{figure*}
\section{\label{sec:Hamil_EQ}Hamiltonian Formalism Using Coordinate Transformation}
The Hamiltonian (total energy) $H_{M}$ of the 1-DOF electrostatic MEMS actuator, driven by a voltage source, neglecting damping, is given by \cite{fargas2005electrostatic,fargas2007resonant}
\begin{equation} \label{eq:MEMS_EX}
H_{M}(x,\dot{x},t) = \frac{1}{2}~m~\dot{x}^2 + \frac{1}{2}~k~x^2 - \frac{1}{2}~\frac{\epsilon_o~A_M~V_M^2(t)}{(g_o - x)}
\end{equation}
The first term represents the kinetic energy with $\dot{x}=\frac{dx}{dt}$ denoting the velocity. The second and third terms represent the potential energy stored in the spring and in the capacitor formed by the top and bottom electrodes respectively. The negative sign in the third term is due to the energy lost by the voltage source in charging the parallel plate capacitor. Now, we employ a coordinate transformation from displacement to charge. Since the electrostatic MEMS actuator resembles a parallel-plate capacitor, the charge $q$ on the electrode can be related to the displacement $x$ of the electrode as
\begin{equation} \label{eq:QX_Relation}
q = \frac{\epsilon_o~A_M~V_M(t)}{(g_o - x)}
\end{equation} 
Therefore, $H_M$ is obtained as a function of charge $q$ as 
\begin{equation} \label{eq:MEMS_EQ}
\begin{split}
H_M(q,\dot{q},t)= \frac{1}{2}~m~\left(\frac{\epsilon_o~A_M~V_M(t)}{q^2}\right)^2~\dot{q}^2 ~+ ~U_M(q,t)
\end{split}
\end{equation}
where $\dot{q}= \frac{dq}{dt}$ is the current. The first term represents the kinetic energy. The second term denotes the potential energy of the spring and the parallel plate variable capacitor, expressed in the charge coordinate as
\begin{equation} \label{eq:MEMS_UQ}
U_M(q,t) = \frac{1}{2}~k~\left(g_o - \frac{\epsilon_o~A_M~V_M(t)}{q}\right)^2 - \frac{q~V_M(t)}{2}
\end{equation} 
Note that Eq. (\ref{eq:MEMS_EQ}) describes the energy of the electrostatic MEMS actuator for \textit{any} form of voltage actuation $V_M(t)$. We would like to reiterate that the expression for energy derived in Ref. \cite{masuduzzaman2014effective} is valid \textit{only} at points of static equilibrium, because, the mapping from displacement to charge used therein is obtained by equating the electrostatic force of attraction between the two electrodes and the mechanical spring restoring force, which is valid only at points of static equilibrium. On the other hand, the mapping Eq. (\ref{eq:QX_Relation}), used to obtain Eq. (\ref{eq:MEMS_EQ}), describes the charge-voltage relationship of a parallel plate capacitor, and is valid for \textit{any} voltage $V_M(t)$. 

\section{\label{sec:Energy_Charge_Landscape}Analysis of Electrostatic MEMS Actuator based on Energy-Charge Landscape}
\subsection{\label{subsec:Static}Static Pull-in}
\begin{figure*}[t]
    \centering
    \includegraphics[scale=1.0]{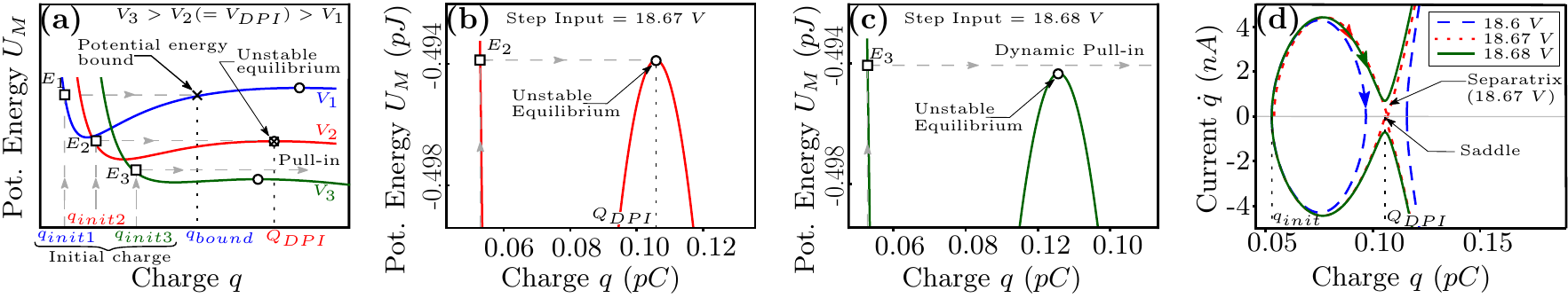}
    \caption{(a) Illustration of dynamic pull-in using potential energy ($U_M$) - charge ($q$) plot. Dynamic pull-in does not occur when initial energy is less than the energy at unstable equilibrium ($V_M<V_{DPI}$). When step-input amplitude $V_M = V_{DPI}$, initial energy equals the energy at unstable equilibrium. Dynamic pull-in occurs for $V_M>V_{DPI}$. Plots for the designed MEMS actuator when (b) $V_M = 18.67 ~V \equiv V_{DPI}$ and (c) $V_M = 18.68 ~V$, depicting dynamic pull-in. (d) Phase portrait for three different step inputs. For $V_M<V_{DPI}$, the closed trajectory implies oscillatory response of the actuator. The voltage corresponding to the separatrix is $V_{DPI}$. The separatrix goes through a saddle point which corresponds to $Q_{DPI}$. Any step-input with $V_M>V_{DPI}$ results in an open trajectory, hence leading to dynamic pull-in.}
    \label{fig:Dynamic_EQ_All}
\end{figure*}
At any given time $t$, let the amplitude of the input voltage be $V_M$. The total energy ($H_M$) as a function of charge ($q$) and current ($\dot{q}$) for an input voltage $V_M = 18 ~V$ is plotted in Fig. \ref{fig:MEMS_3D}(a). In order to find the static equilibria of the system, the time derivatives should be set to zero. Thus, the total energy $H_M$ reduces to the potential energy $U_M$. In Fig. \ref{fig:MEMS_3D}(a), this corresponds to the projection of the total energy on the plane where $\dot{q}=0$. By doing so, we obtain the potential energy ($U_M$) - charge ($q$) landscape for the applied voltage as shown in Fig. \ref{fig:MEMS_3D}(b). The static equilibria correspond to $dU_M/dq = 0$. For each applied voltage, there are two equilibrium charges: stable (local minima with $d^2U_M/dq^2 > 0$) and unstable (local maxima with $d^2U_M/dq^2 < 0$). The stable and unstable equilibrium charges are denoted by the cross ($\times$) and circle ($\circ$) markers respectively. For an input voltage $V_M$, the displacement of the top electrode settles at a position corresponding to the energetically favorable stable equilibrium charge. With increase in $V_M$, the stable and unstable equilibrium charges become more closely spaced in the energy-charge landscape, eventually coinciding with each other. Beyond the static pull-in voltage $V_{SPI}$, there exists no stable equilibrium charge. We define the charge corresponding to this voltage as the static pull-in charge $Q_{SPI}$, as shown in Fig. \ref{fig:MEMS_3D}(b). Thus, beyond $V_{SPI}$, the top electrode snaps down onto the bottom electrode. 

The slope of the potential energy with respect to charge is 
\begin{align} \label{eq:slope_UQ}
\frac{dU_M}{dq} &= k(\epsilon_oA_M)^2V_M \left[\frac{f(q)-V_M}{q^3}\right]\\
\text{with}~~~~~
f(q) &= \left[\frac{g_o}{\epsilon_o A_M} \right]~q - \left[\frac{1}{2k(\epsilon_o A_M)^2} \right]~q^{3}
\end{align}
At static equilibrium, $dU_M/dq=0$. Thus, we obtain the input voltage $V_M$ as a function of the equilibrium charge $q_*$ as
\begin{align} \label{eq:MEMS_QV_basic}
V_M = \left[\frac{g_o}{\epsilon_o A_M} \right]~q_* - \left[\frac{1}{2k(\epsilon_o A_M)^2} \right]~q_*^{3}~\equiv f(q_*)
\end{align}
as shown in Fig. \ref{fig:MEMS_3D}(c). To investigate the stability of the equilibrium charge $q_*$, we obtain 
\begin{equation} \label{eq:sec_deriv}
\frac{d^2U_M}{dq^2}\Bigr|_{\substack{\\q=q_*}} = \frac{k(\epsilon_oA_M)^2V_M}{q_*^{3}}f^\prime(q_*) 
\end{equation}
where $f^\prime(q_*)=\frac{df(q)}{dq}\Bigr|_{\substack{\\q=q_*}}$, is the reciprocal of the slope of the plot in Fig. \ref{fig:MEMS_3D}(c). Thus, from Eq. (\ref{eq:sec_deriv}), we conclude that the equilibrium charge $q_*$ is stable (unstable) when $f^\prime(q_*)$ is positive (negative). The stable and unstable equilibrium charges coincide at $q_*=Q_{SPI}$ when $V_M=V_{SPI}$. 

Using Eq. (\ref{eq:MEMS_QV_basic}) and imposing $d^2U_M/dq^2 = 0$ at pull-in, since pull-in represents an inflection point, we obtain 
\begin{align} \label{eq:V_{SPI}}
V_{SPI}= \sqrt{\frac{8~k~g_o^3}{27~\epsilon_o~A_M}}~;~ Q_{SPI} = \sqrt{\frac{2 ~\epsilon_o ~k ~g_o ~A_M}{3}}
\end{align}
\subsection{Dynamic Pull-in}
For dynamic pull-in, the transient effects due to the applied step-input of amplitude $V_M$ should be considered. The initial conditions $x(0^+)=0$ and $\dot{x}(0^+)=0$ are translated to the charge coordinate as $q(0^+) = q_{init}=(\epsilon_0~A_M~V_M)/g_o$ and $\dot{q}(0^+)=0$, respectively using Eq. (\ref{eq:QX_Relation}). Note that the electrostatic MEMS actuator gets charged to $q_{init}$ instantaneously at $t=0$. This is similar to the case of charging a capacitor in a circuit  without any resistance (see for example Ref. \cite{mac2004network}). As $\dot{q}(0^+)=0$, the total energy reduces to the potential energy and therefore, the initial energy is calculated from Eq. (\ref{eq:MEMS_UQ}) with $q=q_{init}$. Fig. \ref{fig:Dynamic_EQ_All}(a) explains the concept of dynamic pull-in using the \textit{potential energy ($U_M$) - charge ($q$) profile}. When a step-input of amplitude $V_1$ is applied at $t=0$, the initial energy obtained is denoted as $E_1$. The charge on the actuator causes a non-zero acceleration at $t=0$. As a result, the top electrode starts moving, converting potential energy into kinetic energy. However, the displacement of the top electrode is limited by the potential energy bound in the potential energy-charge landscape, as shown in Fig. \ref{fig:Dynamic_EQ_All}(a). This results in an oscillatory response of the actuator in the charge coordinate, similar to the oscillatory response in the displacement coordinate depicted in Fig. \ref{fig:Input_Output}(b)(i). The oscillations are now between the initial charge $q_{init1}$ and the corresponding charge $q_{bound}$ as depicted in Fig. \ref{fig:Dynamic_EQ_All}(a). When the amplitude of the step-input is increased to $V_2$, the initial energy $E_2$ equals the energy at the unstable equilibrium and this input corresponds to the dynamic pull-in voltage $V_{DPI}$. We define the unstable equilibrium charge corresponding to $V_{DPI}$ as the dynamic pull-in charge $Q_{DPI}$. Any further increase in amplitude of the step voltage will result in the initial energy being greater than the energy at the unstable equilibrium. Hence, this will result in dynamic pull-in as depicted for a step-input of amplitude $V_3$, in Fig. \ref{fig:Dynamic_EQ_All}(a). Thus, $V_{DPI}$ and $Q_{DPI}$ are derived using the condition that, at dynamic pull-in voltage, the initial energy is equal to the energy at dynamic pull-in charge; that is, when $V_M = V_{DPI}$, we have $U_M(q = q_{init}) = U_M(q=Q_{DPI})$. Using this and the fact that $Q_{DPI}$ is also an equilibrium charge with $dU_M/dq = 0$ at $Q_{DPI}$, we obtain, 
\begin{align}  \label{eq:V_{DPI}}
V_{DPI}=\sqrt{\frac{k~g_o^3}{4~\epsilon_o~A_M}}~;~ Q_{DPI} = \sqrt{\epsilon_o ~k ~g_o ~A_M}
\end{align}   
Dynamic pull-in can also be visualized using the phase portrait. The phase plane is obtained from the 3D plot shown in Fig. \ref{fig:MEMS_3D}(a), by taking the projection on the plane where total energy is \textit{constant}. This constant is fixed by the initial energy. Each trajectory in the phase plane shows the evolution of a set of initial conditions ($q$ and $\dot{q}$), with time, for an applied step input. The collection of such trajectories for different applied voltages is called the phase portrait as shown in Fig. \ref{fig:Dynamic_EQ_All}(d). Note that the values of the initial charge $q_{init}$ for the three voltages are numerically very close and hence, appear to be the same charge in the phase portrait. For an applied step-input of amplitude 18.6 $V$, the closed trajectory in the phase portrait implies oscillatory response of the actuator. The dynamic pull-in voltage ($V_{DPI}=18.67 ~V$) manifests in the form of a separatrix that separates the behaviour before pull-in and after pull-in. The separatrix goes through a saddle point which corresponds to $Q_{DPI}$. Any step-input with amplitude greater than $V_{DPI}$ ($V_M =$ 18.68 $V$) will result in dynamic pull-in, as depicted by the open trajectory in the phase portrait. 
\begin{figure*}[t]
    \centering
    \includegraphics[scale=1.0]{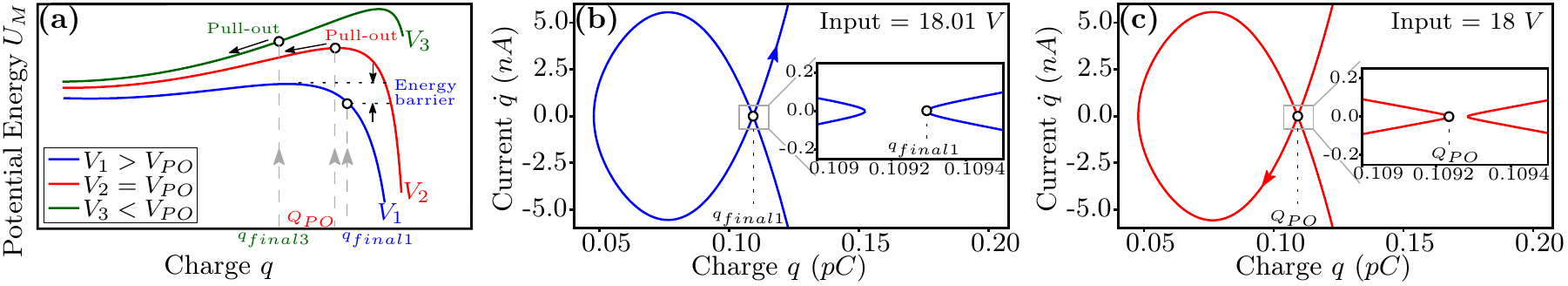}
    \caption{(a) Illustration of pull-out phenomenon using potential energy ($U_M$) - charge ($q$) plot. For input voltage $V_1 > V_{PO}$, the presence of energy barrier at $q_{final1}$ prevents pull-out. For input voltage $V_2 = V_{PO}$, the energy barrier just disappears at $q_{final2} = Q_{PO}$, resulting in pull-out. For input voltage $V_3 < V_{PO}$, the absence of energy barrier at $q_{final3}$ results in pull-out. Phase portrait for step-input with amplitude (b) $18.01~V$ and (c) $18~V$. Release of the top electrode is not achieved when the step-input is reduced to $18.01~V$ as $q_{final1}$ lies on the open trajectory. When the input is reduced to $V_{PO}=18~V$, the corresponding charge $Q_{PO}$ lies on the closed trajectory and hence the top electrode gets released.}
    \label{fig:Release_EQ}
\end{figure*}
\subsection{Pull-out}
After achieving pull-in (static or dynamic), the top electrode has moved a distance of $x_{PO}=g_o - h_s$. In the charge coordinate, using Eq. (\ref{eq:QX_Relation}), this corresponds to a charge $q_{final} = \epsilon_oA_MV_M/h_s$. 
The pull-out phenomenon can be understood using the potential energy ($U_M$) - charge ($q$) plot as illustrated in Fig. \ref{fig:Release_EQ}(a). The top electrode remains attached to the bottom electrode as long as there exists an energy barrier ($dU_M/dq<0$) at charge $q_{final1}$ for an applied voltage $V_1$. At the pull-out voltage $V_{PO}$, the energy barrier disappears ($dU_M/dq = 0$). That is, for $V_2 = V_{PO}$, we have $q_{final2}=Q_{PO}= \epsilon_oA_MV_{PO}/h_s$. Any applied voltage less than $V_{PO}$ also results in pull-out ($dU_M/dq>0$ at $q_{final3}$), as illustrated in Fig. \ref{fig:Release_EQ}(a). Since the slope of the potential energy-charge profile is zero at $Q_{PO}$ for applied voltage $V_M = V_{PO}$, from Eq. (\ref{eq:slope_UQ}), setting $\frac{dU_M}{dq}= 0$ with $q=Q_{PO}$, we derive  
\begin{equation}  \label{eq:V_{PO}}
V_{PO} = \sqrt{\frac{2~k~h_s^2~(g_o - h_s)}{\epsilon_o~A_M}}
\end{equation} 

Pull-out can also be visualized using the phase portrait as shown in Fig. \ref{fig:Release_EQ}(b),(c). Release of the top electrode is not achieved when the step-input is reduced to $18.01~V$ as $q_{final1}$ lies on the open trajectory. When the input is further reduced to $V_{PO}=18~V$, the corresponding charge $q_{final}(V_M=V_{PO})=Q_{PO}$ lies on the closed trajectory and hence the top electrode gets released. The closed trajectory illustrates the sustained oscillatory response of the top electrode, after release, in the absence of damping. 
\begin{table}[b]
\caption{\label{tab:Table3}%
Pull-in and pull-out of an electrostatic MEMS actuator based on energy-charge landscape
}
	\centering
	\begin{adjustbox}{width=\columnwidth,center}
	\begin{tabular}{lll}
		\toprule
		\textbf{Condition}&
		\textbf{Voltage}&
		\textbf{Charge}\\
		\midrule
		\makecell[l]{Static pull-in} & $V_{SPI}=\sqrt{(8~k~g_o^3)/(27~\epsilon_o~A_M)}$ & $Q_{SPI} = \sqrt{2 ~\epsilon_o ~k ~g_o ~A_M/3}$\\
		\makecell[l]{Dynamic pull-in} & $V_{DPI}=\sqrt{(k~g_o^3)/(4~\epsilon_o~A_M)}$ & $Q_{DPI} = \sqrt{\epsilon_o ~k ~g_o ~A_M}$\\
		\makecell[l]{Pull-out} & $V_{PO}=\sqrt{(2~k~h_s^2~(g_o - h_s))/(\epsilon_o~A_M)}$ & $Q_{PO} = \epsilon_o~A_M~V_{PO}/h_s$\\
		\bottomrule
	\end{tabular}
	\end{adjustbox}
\end{table}

Note that the voltage expressions derived above for static pull-in, dynamic pull-in and pull-out using the energy-charge landscape are identical with those derived from the energy-displacement landscape (Table \ref{tab:Table2}). Whether the input voltage is varied slowly as in the case of static input, or the input voltage is varied suddenly as in the case of dynamic input, the pull-out voltage is the same. This is because the actuator remains at $x = g_o - h_s$ until the input voltage $V_M$ is reduced to $V_{PO}$, be it slowly or suddenly, leading to disappearance of the energy barrier. Thus it is the disappearance of the energy barrier that decides the pull-out rather than the manner by which the input voltage is varied. Contrast this with the situation during pull-in: the electrostatic MEMS actuator can pull-in either if the system does not see a barrier or the system has sufficient energy to surmount the barrier. The former case happens for slowly varying input $V_M$ (static pull-in) or step excitation (dynamic pull-in) with $V_M > V_{SPI}$. The latter case happens only for step excitation with $V_M > V_{DPI}$ and $V_M<V_{SPI}$ (that is, $V_{DPI}< V_M < V_{SPI}$). Hence, it is the nature of the energy landscape that brings out the above described contrast between pull-in and pull-out. 
The expressions for the voltage and charge, derived using the proposed framework, are summarized in Table \ref{tab:Table3}.
\section{Impact of damping}
\begin{figure*}[t]
    \centering
    \includegraphics[scale=1.0]{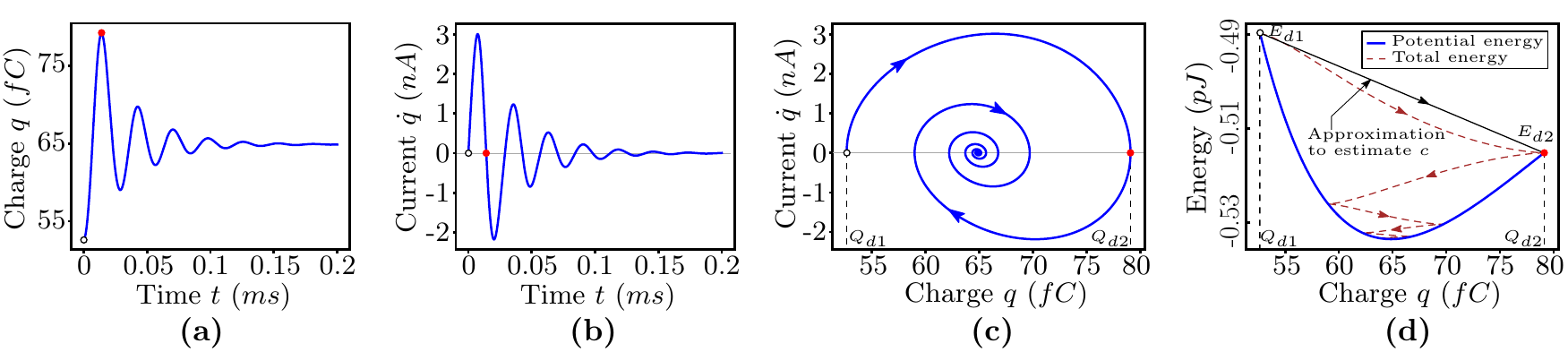}
    \caption{Plots for step input actuation with $V_M=18.6~V$ and $\zeta=0.1$. (a),(b) Transient response of the charge and current. (c) Phase portrait depicting the decaying oscillations of the charge and current. (d) Potential energy - charge plot, along with the time evolution of the total energy. Damping coefficient $c$ is estimated based on the approximation in Ref. \cite{casals2014analytical}. The estimated value of $c$ is $3.5 \times 10^{-7}~Ns/m$ (exact value of $c$ is $2.8 \times 10^{-7}~Ns/m$).}
    \label{fig:Param}
\end{figure*}
We have neglected damping in our analysis so far. We now examine the impact of damping by including the damping coefficient $c$ in the 1-DOF model in Fig. \ref{fig:MEMS_1DOF}. Here, $c$ represents an effective value accounting for various damping mechanisms in MEMS \cite{younis2011mems}. We calculate the damping coefficient as $c = 2 m \omega_o \zeta$, where $\omega_o=\sqrt{k/m}$ is the natural frequency and $\zeta$ is the damping ratio of the mechanical cantilever structure \cite{younis2011mems}. The dynamic pull-in voltage is influenced by damping \cite{shekhar2012switching}. With increase in damping, the dynamic pull-in voltage approaches $V_{SPI}$. The dynamic response, before pull-in, decays with time and the actuator settles at the static equilibrium displacement corresponding to the magnitude of the applied step input. The transient response in the charge coordinate also shows a similar behaviour. For example, Fig. \ref{fig:Param}(a),(b) show the transient charge $q(t)$ and current $\dot{q}(t)$ plotted for $\zeta = 0.1$ and $V_M=18.6~V$. These plots have been obtained by numerically solving the MEMS dynamics in the charge coordinate. The corresponding phase portrait is plotted in Fig. \ref{fig:Param}(c), where the trajectory is an inward spiral (unlike the closed trajectory for the undamped case in Fig. \ref{fig:Dynamic_EQ_All}(d)). Using Eq. (\ref{eq:MEMS_EQ}) and Eq. (\ref{eq:MEMS_UQ}), we also plot the total energy and the potential energy, as a function of charge, as shown in Fig. \ref{fig:Param}(d). The total energy of the system evolves with time, as depicted by its trajectory and finally, the system settles at the static equilibrium charge. 

Electrical measurement techniques for estimation of various MEMS parameters are common and are widely reported in literature  \cite{bhat2007parameter,basu2007estimation,hu2005nonlinear,osterberg1997m,clark2004practical}.
We now propose a procedure to estimate various parameters using the energy-charge landscape, based on electrical measurements.  For instance, an electrical measurement set-up (such as, in Ref. \cite{bhat2007parameter}) could be used to measure the transient current $\dot{q}(t)$, for a step-input excitation with $V_M$ less than the pull-in voltage. The transient charge $q(t)$ can then be obtained by integrating $\dot{q}(t)$. With the help of the transient response and energy plots, we can estimate parameters such as displacement, velocity, air-gap, spring constant and damping coefficient. Velocity is estimated as $\dot{x}(t) = \epsilon_o A_M V_M(t) \dot{q}(t)/q^2(t)$. Displacement $x(t)$ can be obtained by integrating $\dot{x}(t)$. Note that the final steady state value of the charge in the transient response in Fig. \ref{fig:Param}(a), corresponds to the stable equilibrium charge of the static response. This allows us to estimate air-gap $g_o$ and spring constant $k$ from the equilibrium charge-voltage relationship, given by Eq. (\ref{eq:MEMS_QV_basic}). Let the stable equilibrium (steady state) charges be denoted as $Q_a$ and $Q_b$ for two different step inputs of amplitude $V_a$ and $V_b$ respectively ($V_a$,$V_b$ less than the dynamic pull-in voltage). Using Eq. (\ref{eq:MEMS_QV_basic}), we propose the estimation of $g_o$ and $k$ as
\begin{equation} \label{eq:param_gok}
g_o \thickmuskip=3.1mu = \frac{A_M \epsilon_o (V_b Q_a^3 - V_a Q_b^3)}{Q_a Q_b (Q_a^2 - Q_b^2)};~k \thickmuskip=3.1mu = \frac{Q_a Q_b (Q_a^2 - Q_b^2)}{2 A_M^2 \epsilon_o^2 (Q_a V_b - Q_b V_a)}
\end{equation} 
Any point on the potential energy plot in Fig. \ref{fig:Param}(d) corresponds to zero kinetic energy, implying $\dot{q}=0$, according to Eq. (\ref{eq:MEMS_EQ}). Thus, we can obtain the potential energies $E_{d1}$ and $E_{d2}$ in Fig. \ref{fig:Param}(d), corresponding to the two consecutive charges $Q_{d1}$ and $Q_{d2}$ on the phase portrait, where $\dot{q}=0$ (see Fig. \ref{fig:Param}(c)). The energy dissipated during this time interval can be calculated as $\Delta E_d = E_{d1}-E_{d2}$. From the estimated $x(t)$ and $\dot{x}(t)$, the distance travelled during this time interval, $x_{d}$, and the average velocity for traversing this distance, $v_{avg}$, can also be calculated. Based on Ref. \cite{casals2014analytical}, we propose to estimate an approximate value of the damping coefficient as $c \approx \Delta E_d/(x_d v_{avg})$. Based on our simulation, we find this approximation to give a reasonable estimate of $c$ (within $50\%$ of the actual value) for $\zeta$ in the range $0$ to $0.55$. Note that the value of mass $m$ is not needed to determine $E_{d1}$ and $E_{d2}$. The above described technique could be an alternative to the other  methods available \cite{gad2001mems} for the measurement of these parameters.
\begin{figure*}[t]
    \centering
    \includegraphics[scale=1.0]{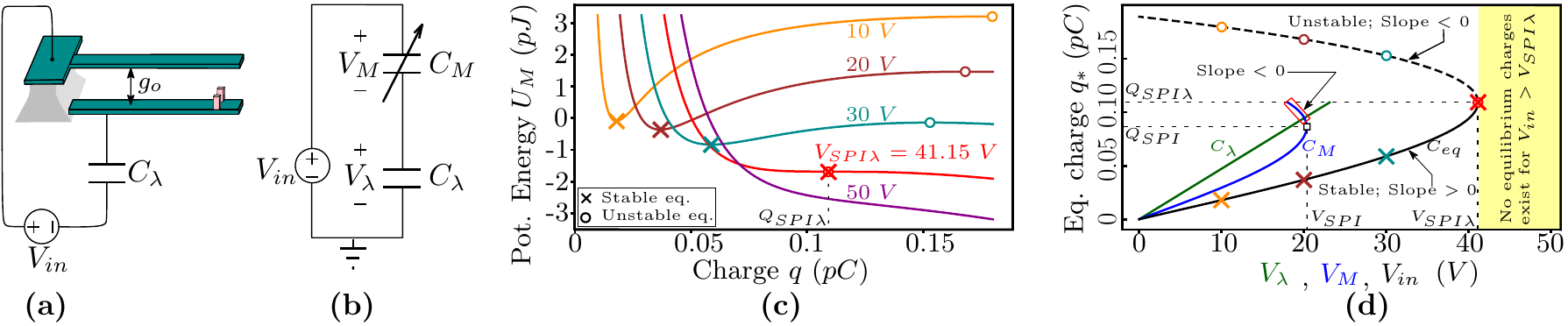}
    \caption{(a) Schematic of the feedback capacitor-MEMS actuator, as proposed in Ref. \cite{seeger1997stabilization}. (b) Equivalent circuit representation: $C_{\lambda}$ represents the series feedback capacitance and $C_M$ denotes the variable capacitance of the MEMS actuator. (c) Potential energy ($U_{\lambda}$) - charge ($q$) plot for different input voltages. The static pull-in charge and static pull-in voltage are $Q_{SPI\lambda}$ and $V_{SPI\lambda}$ respectively. (d) Equilibrium charge-voltage characteristics. Though the equilibrium charge ($q_*$) - $V_M$ plot has a region with negative slope, the actuator is stabilized in this region by the feedback capacitor. Therefore, the static pull-in charge (voltage) is increased from $Q_{SPI}$ ($V_{SPI}$) in the standalone MEMS actuator, to $Q_{SPI\lambda}$ ($V_{SPI\lambda}$) in the feedback capacitor-MEMS actuator. Increase in the static pull-in charge results in the extension of the travel range.}
    \label{fig:Feedback_Cap_MEMS}
\end{figure*}
\section{Scope and Limitations}
We have used a 1-DOF lumped parameter model that neglects the transverse deflection of the cantilever along its length \cite{ijntema1992static,younis2011mems}. We have also neglected the effect of the fringing field capacitance in our analysis. The effect of the fringing field capacitance can be captured by a modified MEMS capacitance expression available in Ref. \cite{nemirovsky2001methodology}. We have ignored the effect of surface forces during collision between the top electrode and the stopper. The methodology presented, however, can be extended to include the surface forces as well, since these surface forces manifest as additional terms in the Hamiltonian \cite{granaldi2006dynamic}. Additionally, a new mapping function has to be formulated for other geometries and systems. As long as the MEMS structure is capacitive, the proposed framework can be used, with the new mapping function describing the charging equation of the MEMS capacitance. An example for this is provided in the case study presented in the next section. 
\section{Case Study: Feedback Capacitor-MEMS Actuator System}
We consider the case of a feedback capacitor connected in series with an electrostatic MEMS actuator, excited by a voltage source \cite{seeger1997stabilization}, as shown in Fig. \ref{fig:Feedback_Cap_MEMS}(a),(b). Here, $C_M$ denotes the variable capacitance of the MEMS actuator and $C_{\lambda}$ denotes the fixed feedback capacitance. An extension of the travel range for static input, beyond the conventional travel range of one-third of the air-gap, is proposed in Ref. \cite{seeger1997stabilization}, using the energy-displacement landscape. An experimental validation of this extension of the travel range using the capacitive feedback is presented in Ref. \cite{chan2000electrostatic}.  

We now analyze this feedback capacitor-MEMS actuator system using the energy-charge landscape. The variable displacement is defined \emph{only} for the MEMS actuator (as the feedback capacitor does not have any movable part). On the other hand, \emph{both} the capacitors share the same charge as they are connected in series. Hence, it is convenient to use charge for the analysis. This provides an advantage over the energy-displacement approach as we can now look at the state of the individual components of the system separately, with charge being the common variable. Thus, we analyze the operation of the actuator by plotting the charge-voltage characteristics for the overall system and also for the individual capacitances: MEMS capacitance $C_M$ and feedback capacitance $C_{\lambda}$. 

We define $\lambda$ as the ratio of zero-bias MEMS capacitance $C_o=\epsilon_o A_M/g_o$ to the feedback capacitance $C_{\lambda}$. Since $C_{\lambda}$ and $C_M$ are in series, the equivalent capacitance $C_{eq} = C_{\lambda}~C_M/(C_{\lambda} + C_M)$. Substituting $C_M=\epsilon_o~A_M/(g_o-x)$ and $C_{\lambda}=C_o/\lambda$, we obtain $C_{eq} = (\epsilon_o~A_M)/(g_o(\lambda+1)-x)$. This implies that the effective electrical air-gap of the actuator is now $g_o(\lambda + 1)$. 
The feedback capacitor-MEMS actuator is excited by a voltage source $V_{in}$. Therefore, we obtain the potential energy of the system as 
\begin{equation} \label{eq:energy_feedback_MEMS}
U_{\lambda}(x,t) = \frac{1}{2}~k~x^2 - \frac{1}{2}~\frac{\epsilon_o~A_M~V_{in}^2(t)}{g_o(\lambda+1)-x}
\end{equation}
The voltage across MEMS actuator $V_M(t)$ is related to the input voltage $V_{in}(t)$ as
\begin{equation} \label{eq:vol_divison}
V_M(t) = V_{in}(t)\frac{C_{\lambda}}{C_{\lambda}+C_M} = \frac{V_{in}(t)}{1+[\lambda ~g_o/(g_o-x)]}
\end{equation}
Since $C_{\lambda}$ and $C_M$ are in series, the charge $q$ remains the same on both the capacitors. Substituting Eq. (\ref{eq:vol_divison}) in our original mapping function, given by Eq. (\ref{eq:QX_Relation}), we obtain the relation between charge $q$ and displacement $x$ for the feedback capacitor-MEMS actuator as
\begin{equation} \label{eq:mapping_feedback_disp}
q = \frac{\epsilon_o~A_M~V_{in}(t)}{g_o(\lambda+1)-x}~;~ x = g_o(\lambda+1) - \frac{\epsilon_o~A_M~V_{in}(t)}{q}
\end{equation}  
Substituting Eq. (\ref{eq:mapping_feedback_disp}) in Eq. (\ref{eq:energy_feedback_MEMS}), we obtain the potential energy of the system as a function of charge $q$ as
\begin{equation}
U_{\lambda}(q,t) = \frac{k}{2}\left[g_o(\lambda+1) - \frac{\epsilon_o A_M V_{in}(t)}{q}\right]^2 - \frac{q}{2}V_{in}(t)
\end{equation} 
The values of the MEMS actuator parameters are the same as in Table \ref{tab:Table1}. We choose $\lambda$ so that the entire distance $g_o-h_s$ equals the extended static travel range $X_{SPI\lambda}$. This ensures stable operation of the actuator over the full range and thus, eliminates the static pull-in instability. Therefore, $X_{SPI\lambda} = g_o(\lambda+1)/3 =  g_o-h_s$. Substituting the values of $g_o$ and $h_s$, we get $\lambda=0.6$. Please note that we could instead choose $\lambda$ such that the dynamic pull-in displacement equals the entire distance $g_o - h_s$. However, in order to compare our results with Ref. \cite{seeger1997stabilization}, we only consider the static case here. 

Fig. \ref{fig:Feedback_Cap_MEMS}(c) shows the potential energy ($U_{\lambda}$) - charge ($q$) plot of the feedback capacitor-MEMS actuator, for different input voltage $V_{in}$. As in the case of the standalone MEMS actuator in Fig. \ref{fig:MEMS_3D}(b), the stable and unstable equilibrium charges coincide, when $V_{in}$ equals the static pull-in voltage of the feedback capacitor-MEMS actuator $V_{SPI\lambda}$. The corresponding charge is the static pull-in charge of the feedback capacitor-MEMS actuator $Q_{SPI\lambda}$, as shown in Fig. \ref{fig:Feedback_Cap_MEMS}(c).

At static equilibrium, $\frac{dU_{\lambda}}{dq}=0$. Therefore, we obtain the equilibrium charge ($q_*$) - input voltage ($V_{in}$) relation for the feedback capacitor-MEMS actuator as
\begin{equation} \label{eq:FMEMS_QV}
V_{in} = \left[\frac{g_o(\lambda+1)}{\epsilon_o A_M} \right]~q_* - \left[\frac{1}{2k(\epsilon_o A_M)^2} \right]~q_*^{3}
\end{equation}

Fig. \ref{fig:Feedback_Cap_MEMS}(d) shows the equilibrium charge ($q_*$) - input voltage ($V_{in}$) plot of the feedback capacitor-MEMS actuator, obtained using Eq. (\ref{eq:FMEMS_QV}). The feedback capacitor-MEMS actuator is stable in the region with positive slope, as shown, with $V_{in}\leq V_{SPI\lambda}$. This is similar to the equilibrium charge-voltage plot of the standalone MEMS actuator in Fig. \ref{fig:MEMS_3D}(c). However, note that the static pull-in charge and static pull-in voltage in the feedback capacitor-MEMS actuator are larger than their corresponding counterparts in the standalone MEMS actuator. This increase in the static pull-in charge results in an extension of the travel range. Substituting $V_{in}=V_{SPI\lambda}$ and $q=Q_{SPI\lambda}$ in Eq. (\ref{eq:mapping_feedback_disp}), we confirm that the travel range of the feedback capacitor-MEMS actuator $X_{SPI\lambda}= 1.6~\mu m \equiv g_o-h_s$, thereby eliminating the static pull-in instability (travel range of the  standalone MEMS actuator $X_{SPI}$ is $1~\mu m$). 

The energy-charge landscape approach enables us to plot the equilibrium charge-voltage characteristics of the individual capacitances: $C_M$ and $C_{\lambda}$, separately. For an applied input voltage $V_{in}$ and the corresponding \emph{stable} equilibrium charge, the voltage across MEMS capacitor $V_M$ is obtained using Eq. (\ref{eq:vol_divison}) and Eq. (\ref{eq:mapping_feedback_disp}). Also, the voltage across the feedback capacitor $V_{\lambda}$ is obtained as $V_{in} - V_M$. Thus, in Fig. \ref{fig:Feedback_Cap_MEMS}(d), we  also show the equilibrium charge ($q_*$) - $V_{\lambda}$ plot of the feedback capacitance $C_{\lambda}$ and equilibrium charge ($q_*$) - $V_M$ plot of the MEMS capacitance $C_M$. Note that the stability of the overall system is determined by the slope of the $q_*$-$V_{in}$ plot (stable when slope is positive). The $q_*$-$V_{\lambda}$ plot is a straight line as $C_\lambda$ is a fixed capacitor. The $q_*$-$V_M$ plot includes a region with negative slope where the actuator is stable. The stability of the actuator in this region is due to the capacitive feedback provided by $C_{\lambda}$. This is unlike the case in the standalone MEMS actuator, where the actuator is unstable in the region with negative slope in its $q_*$-$V_{M}$ plot [see Fig. \ref{fig:MEMS_3D}(c)]. Thus, the improvement in the stability and the extension of the travel range are conveniently explained by the equilibrium charge-voltage plots, derived from the energy-charge landscape.  
\section{\label{sec:Conclusion}Conclusion}
We have presented a unified framework to analyze the statics and dynamics of an electrostatic MEMS actuator from its energy-charge landscape. The proposed method employs coordinate transformation from the conventional displacement coordinate to the charge coordinate. This coordinate transformation is used in the Hamiltonian formalism to obtain the energy-charge relationship. The expressions for the voltage and charge, derived using the proposed framework, are summarized in Table \ref{tab:Table3}. The voltage expressions derived using energy-charge relationship (Table \ref{tab:Table3}) are identical with those derived using the conventional energy-displacement relationship (Table \ref{tab:Table2}). The impact of damping is also examined using the energy-charge method. The proposed framework will aid in the design and analysis of electrostatic MEMS devices. A case study, considering a feedback capacitor-MEMS actuator system, is also presented, to illustrate the convenience of  using the proposed energy-charge landscape in the design and analysis. 
\section*{acknowledgment}
The authors thank Prof. G. K. Ananthasuresh for useful inputs.
\bibliographystyle{IEEEtran}
%\nocite{*}
\bibliography{TED_Reference}
\end{document}